\documentclass[conference]{IEEEtran}
\IEEEoverridecommandlockouts
\usepackage{cite}
\usepackage{pifont}
\usepackage{amsmath,amssymb,amsfonts}
\usepackage{graphicx}
\usepackage{textcomp}
\usepackage{xcolor}
\usepackage{subfig} 
\usepackage{booktabs}
\usepackage{booktabs}
\usepackage{bm}
\usepackage{enumitem}
\usepackage[flushleft]{threeparttable}
\usepackage{makecell} 
\usepackage{adjustbox}

\usepackage{algorithm}
\usepackage{algorithmicx}
\usepackage{algpseudocode}
\usepackage{subcaption}

\usepackage[a4paper, total={184mm,239mm}]{geometry}
\def\BibTeX{{\rm B\kern-.05em{\sc i\kern-.025em b}\kern-.08em
    T\kern-.1667em\lower.7ex\hbox{E}\kern-.125emX}}
\begin{document}

\title{\LARGE \bf A Unified Open-Set Framework for Scalable PUF-Based Authentication of Heterogeneous IoT Devices}

\author{\IEEEauthorblockN{
    Xin Wang\textsuperscript{1},
    Peichun Hua\textsuperscript{1},
    Chip Hong Chang\textsuperscript{2},
    Wenye Liu\textsuperscript{3} and
    Yue Zheng\textsuperscript{1}
}
\IEEEauthorblockA{
    \textsuperscript{1}The Chinese University of Hong Kong (Shenzhen), Shenzhen, Guangdong, China \\
    \textsuperscript{2}Nanyang Technological University, Singapore\\
    \textsuperscript{3}Independent Researcher \\
    \{xinwang3, peichunhua\}@link.cuhk.edu.cn,
    echchang@ntu.edu.sg, wenye.liu@ieee.org, zhengyue@cuhk.edu.cn
}
\thanks{This work has been submitted to the IEEE for possible publication. Copyright may be transferred without notice, after which this version may no longer be accessible.}
}

\maketitle

\begin{abstract}
As modern cyber systems scale to include large populations of heterogeneous IoT devices, securing them against impersonation and forgery is a critical cybersecurity challenge. Physical Unclonable Functions (PUFs) offer a lightweight, hardware-rooted trust anchor for IoT security. However, different PUF architectures possess distinct challenge-response spaces and raw response reliabilities, making existing authentication protocols PUF-type-specific. To bridge this interoperability bottleneck, this paper proposes a scalable, helper-data-free, open-set PUF authentication framework that leverages an OpenGAN-based classifier to manage heterogeneous fleets of IoT devices. Our method addresses the limitations of traditional database-centric and digital-twin modeling methods by encoding raw responses from diverse PUF types, including strong, weak and hybrid PUFs, into a unified image representation. This enables robust, single-pass classification and impostor rejection. We integrate the classifier into a generic protocol employing hybrid encryption and Bloom filter-based replay detection. Evaluated across four different types of noisy PUF data (Arbiter, SRAM, DRAM, and heterogeneous PUFs), our framework achieves 100\% closed-set accuracy and near-zero open-set error rates with up to 45 devices, a significant improvement over the 3–5 devices in prior classification-based approaches. Prototyped on a Raspberry Pi, our framework completes one authentication cycle within 0.67 s, approximately $30\times$ faster than the state-of-the-art open-set baselines.
\end{abstract}

\begin{IEEEkeywords}
IoT device authentication, physical unclonable function, open-set classification, heterogeneous IoT network.
\end{IEEEkeywords}

\section{Introduction}
In an era of increasing cyber threats, ensuring the security and resilience of large-scale cyber systems is paramount. As Internet of Things (IoT) deployments scale to networks comprising heterogeneous edge devices from diverse manufacturers, securing devices against sophisticated impersonation and replay attacks has emerged as a critical cybersecurity challenge~\cite{more2023identity, cirne2024hardware, al2023iotpuf}. Physical Unclonable Function (PUF)~\cite{chang2017retrospective, pappu2002physical} offers a promising threat prevention mechanism by exploiting inherent manufacturing process variations to create unique hardware ``fingerprints". When stimulated by specific input \emph{challenges}, a PUF generates device-unique \emph{responses}. These mappings, known as challenge–response pairs (CRPs), serve as the foundation for identity verification. Due to their unclonable nature and low-cost footprint, PUFs have been widely adopted for lightweight IoT device authentication and on-demand key generation~\cite{al2023iotpuf}. 

\begin{figure}[t!]
    \centering
    \includegraphics[width=1\linewidth]{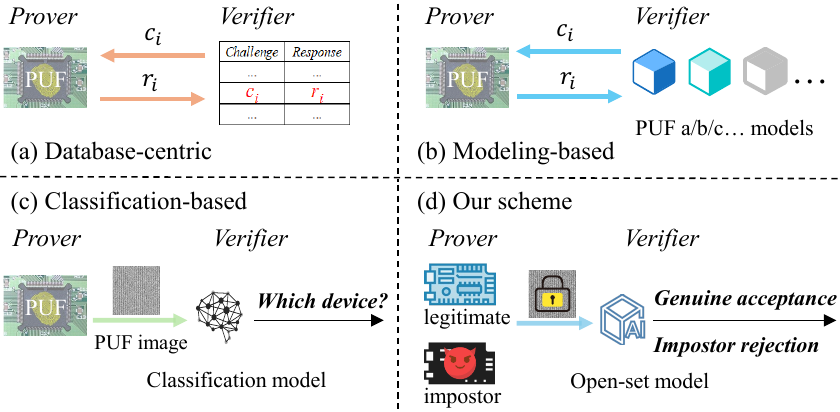}
    \caption{Comparison of PUF-based device authentication paradigms for IoT devices and overview of our open-set authentication framework.}
    \label{fig:comparison}
\end{figure}

Despite their theoretical promise, engineering PUF-based defenses into practical, real-world cyber systems remains a formidable task. Devices in real-world deployments are often equipped with vastly different PUF architectures, broadly categorized as ``weak" or ``strong" based on their available CRP space \cite{gao2020PUFne,gassend2003physical, ulrich2010modeling}. Strong PUFs leverage an expansive CRP space to facilitate unique challenge-response dialogues directly with a server, enabling lightweight IoT device authentication. In contrast, while weak PUFs are highly effective for generating stable, unique cryptographic keys, their limited CRP space prevents them from supporting direct, repeated challenge-response exchanges. Consequently, weak PUFs are typically authenticated via key-based protocols, where they serve as a hardware Root of Trust to derive the secret keys required for traditional cryptographic schemes. Due to the fundamental differences in their operational logic and security models, no single, universally adopted authentication protocol currently exists that seamlessly unifies both weak and strong PUFs.

Challenge-response authentication schemes for strong PUFs typically rely on either static databases or digital-twin modeling (see Fig.~\ref{fig:comparison}). Database-centric schemes~\cite{yu2010secure, muelich2017ecc} require the verifier (typically an authentication server) to pre-collect and store a large set of CRPs for each enrolled device, exhausting each CRP after a single use to prevent replay attacks. This CRP pool must be provisioned carefully to meet long-term authentication needs while avoiding costly and complex re-enrollment procedures. A traditional database for a strong PUF would theoretically require petabytes of storage to house trillions of possible pairs. Modeling-based schemes~\cite{pour2020puf, majzoobi2012slender, zalivaka2018reliable} address this exhaustion problem by training a predictive model (a ``digital twin") for each device. However, securely storing the massive CRP databases or independently training and maintaining individual machine learning (ML) models introduces severe security and resource management challenges as the number of devices increases. Furthermore, since both paradigms rely on an exponentially large challenge space,  ML modeling attacks are a severe threat. Countermeasures against these attacks inevitably introduce structural and protocol diversity, making it infeasible to bridge the heterogeneity gap, even within the strong PUF community.

Edge devices often operate under fluctuating environmental conditions, causing PUFs to produce noisy responses. Error Correction Codes (ECC) are widely adopted to guarantee resilient authentication; however, since different PUFs exhibit distinct raw data distributions and error behaviors, they require highly tailored ECC configurations. Attempting to force a ``one-size-fits-all" error correction protocol across heterogeneous devices imposes a severe design margin penalty, introducing three critical bottlenecks: (1) Security vulnerabilities: ECC requires the generation and storage of helper data to reconstruct keys from noisy responses. Publicly accessible helper data has been demonstrated to leak sensitive PUF information, exposing systems to sophisticated helper data attacks~\cite{becker2017robust, strieder2021machine}. (2) System inefficiency: assuring resilience across diverse PUFs demands the deployment of aggressive, worst-case ECC protocols, which severely degrades overall system efficiency. (3) Hardware and energy costs: BCH decoding imposes prohibitive overheads on IoT devices, and the Berlekamp-Massey algorithm requires $O(t^2)$ operations \cite{massey1969bch, lin2004error} to identify the error-locator polynomial, while root-finding adds $O(n \times t)$  complexity. Consequently, hardware and energy costs scale quadratically with error-correcting capability $t$ \cite{park2018area, nabipour2024trends}. These challenges are further complicated by the fact that weak and strong PUFs utilize helper data differently. In weak PUFs, the primary goal is to reconstruct a bit-perfect cryptographic key upon power-up; even a single bit error will cause the final key to fail completely. Recently, Weak-PUF-assisted strong PUFs~\cite{liu2022weak} have attempted to combine the stability of weak PUFs with the expansive challenge space of strong PUFs. For the device to be successfully authenticated, the key generated by its weak PUF must be entirely error-free. Conversely, strong PUFs frequently use helper data for error avoidance rather than just correction, e.g., instructing the system to ignore specific physical components known to be unstable or highly sensitive to noise by ``dark bit" masking \cite{hiller2016cherry, he2023asch}.

Recently, \emph{classification-based} schemes have emerged as an alternative for weak PUF authentication. By treating responses from a device's 2D PUF cell array, generated under varying environmental factors, as biometric patterns, these schemes shift verification from strict bit-for-bit matching to a computer-vision-based ML problem, enabling a single classifier to learn spatial features to distinguish these 2D patterns across devices within a group. This ``helper-less" model effectively alleviates the denoising burden on low-power devices and overcomes the structural limitations that typically hinder weak PUFs from supporting secure, repeated authentication. While promising, this paradigm remains in a nascent stage; existing works each possess limitations regarding the interoperability bottleneck, as discussed in Sec. \ref{sec:related works}. In this paper, we propose a \textbf{scalable, helper-less, open-set classification-based authentication} framework designed to manage heterogeneous IoT fleets. We leverage the OpenGAN~\cite{Opengan2021} architecture to \emph{actively learn} the decision boundary between known and unknown devices' latent feature distributions, enabling single-pass open-set decisions with minimal computational overhead. To bridge the heterogeneous hardware gap, our framework transforms raw response data from diverse PUF types (e.g., Arbiter, SRAM, DRAM) into a unified image representation. This allows a single classifier to authenticate devices with vastly different PUF architectures, eliminating the need for PUF-specific protocols. Finally, we deploy a \textbf{generic secure authentication protocol} featuring hybrid encryption and a Bloom filter~\cite{bloom1970} for lightweight replay detection, demonstrating the end-to-end practical viability of our system. Our main contributions are summarized as follows:
 
\begin{itemize}[leftmargin=*]
    \item \textbf{Open-Set Formulation:} We utilize an OpenGAN-based classifier to actively learn a rejection boundary without requiring real outlier samples during training. This allows the system to identify enrolled devices while robustly rejecting unknown impostors in a single forward inference.

    \item \textbf{PUF-Agnostic Unification:} We unify \emph{heterogeneous architectures}, including both delay-based strong PUFs and memory-based weak PUFs, into a generic framework. By encoding raw responses into grayscale images, we eliminate the restrictive requirement for uniform PUF hardware across the network. 
    
    \item \textbf{Extensive Empirical Validation:} We evaluate our framework across four distinct PUF data settings: Arbiter (25 devices), SRAM (40 devices)~\cite{wilde2017large}, noisy DRAM (3 devices)~\cite{millwood2023dpan}, and hybrid (45 devices) PUF response datasets. Our system achieves 100\% closed-set accuracy and near-zero open-set error rates (FAR/FRR) across all scenarios. 
    
    \item \textbf{Hardware Prototyping:} We demonstrate efficiency by deploying our framework on a Raspberry Pi, achieving end-to-end authentication in just 0.67\,s. This outperforms state-of-the-art baselines~\cite{yue2020dram, millwood2023dpan, BCNN2024, joshi2025error} and proves our universal approach is fundamentally compatible with standard cryptographic pipelines. 
\end{itemize}

\section{Related Works and Preliminaries}
\subsection{Classification-based PUF Device Authentication} 
\label{sec:related works}
Several classification-based PUF authentication schemes for IoT devices have been proposed to replace conventional database-centric and modeling-based authentication~\cite{yu2010secure, muelich2017ecc} with ML algorithms. Existing works by Karimian et al.~\cite{yue2020dram}, Millwood et al.~\cite{millwood2023dpan}, and Joshi et al.~\cite{joshi2025error} treat the memory cells' states as grayscale images and use CNNs for enrolled device identification. While these approaches demonstrate high recognition accuracy for small groups (typically $\le 5$ devices), they operate under a closed-set assumption, where every device encountered at test time is assumed to be enrolled. Since standard classifiers utilize a softmax layer that distributes 100\% of the probability mass across known classes, they are fundamentally incapable of rejecting forged or unknown devices, creating a critical security vulnerability.

Recent efforts to address this open-world challenge, such as the work by Mexis et al.~\cite{BCNN2024}, employ Bayesian Neural Networks (BNNs) to estimate the prediction uncertainty. While BNNs enable imposter rejection to some extent, they rely on post-hoc ranking rules and softmax confidence thresholds rather than explicitly learning a known-versus-unknown boundary. This makes the system fragile: legitimate devices under environmental noise may be erroneously rejected as their confidence scores fluctuate. Indeed, our reproduction using a public noisy DRAM PUF dataset reveals that the BNN approach suffers from a high FRR of 10.64\% (see Table \ref{tab:comparison}). Furthermore, BNNs require extensive Monte Carlo weight sampling and repeated inference, resulting in unacceptable authentication latency exceeding 20 seconds per query. 

In summary, existing classification-based schemes remain unsuitable for realistic IoT networks due to high latency, limited scalability, poor heterogeneous hardware compatibility, and a lack of robust impostor rejection. These limitations drive us to develop a scalable authentication framework that explicitly learns the decision boundary between enrolled and unknown device distributions, ensuring both robust security and practical efficiency.

\begin{figure*}[!ht]
    \centering
    \includegraphics[width=1\linewidth]{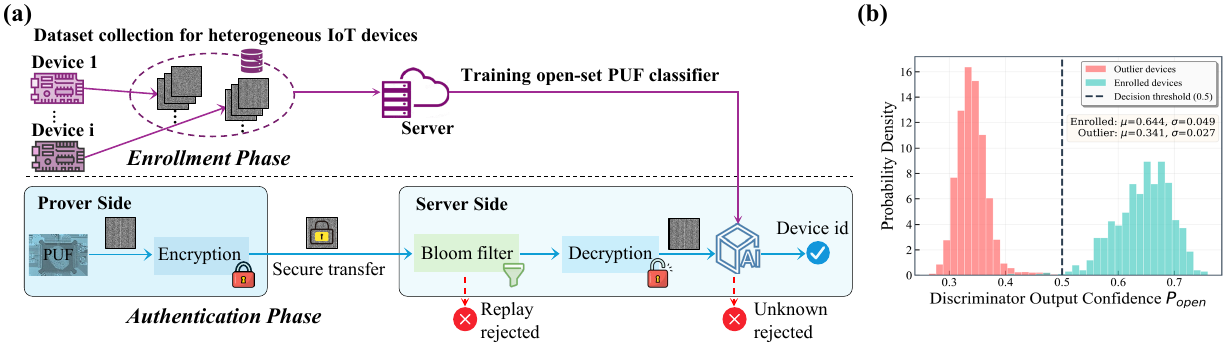}
    \caption{(a) The overall scheme of our proposed framework; (b) Distribution of discriminator outputs on validation set.}
    \label{fig:overall-open-puf-a}
\end{figure*}

\subsection{OpenGAN} 
OpenGAN~\cite{Opengan2021} is a lightweight framework for open-set classification. It augments a $\textit{K}$-way classifier with a generator–discriminator pair $(G, D)$ for open-set discrimination. Let $\mathbf{f} \in \mathbb{R}^d$ denote the pre-logit feature extracted by the $\textit{K}$-way classifier. During training, $G$ synthesizes pseudo open-set features $\tilde{\mathbf{f}} = G(\mathbf{z})$, while $D$ learns to distinguish known features $\mathbf{f}$ from unknown $\tilde{\mathbf{f}}$. At inference, the classifier yields a label $\hat{y}$ and feature $\mathbf{f}$; the discriminator output $P_{\text{open}} = D(\mathbf{f})  \in [0,1]$ estimates the likelihood of the sample belonging to the known set. The prediction $\hat{y}$ is accepted only if $P_{\text{open}} > \tau$, where $\tau$ is a confidence threshold. This enables single-pass open-set rejection without repeated stochastic inference. The appropriate $\tau$ can be determined through the discriminator's confidence score distribution over the validation set, as shown in Fig. \ref{fig:overall-open-puf-a}(b), to avoid over-rejecting or incorrectly accepting.

\subsection{Bloom Filter}
A Bloom filter~\cite{bloom1970} is a probabilistic set-membership structure represented by an $m$-bit vector $\mathbf{B}$ and $k$ hash functions $\{h_i\}_{i=1}^{k}$.
For an element $x$, all indices ${h_i(x)}$ are set to 1 upon insertion. A query returns \textit{true} if all bits at those positions are 1, indicating $x$ has ``possibly'' been seen.
Bloom filters guarantee \textit{no} false negatives but allow false positives with a probability:
\begin{equation}
p = \left(1 - e^{-kn/m}\right)^k
\end{equation}
where $n$ is the number of inserted elements. 

Given a target $p$, the optimal parameters~\cite{tarkoma2011theory} are:
\begin{equation}
m = -\frac{n\ln p}{(\ln 2)^2} \enspace \text{and} \enspace k = \frac{m}{n}\ln 2
\end{equation}

Their low space and time complexity ($O(k)$) makes Bloom filters ideal for fast replay detection in lightweight authentication.

\section{System Model and Threat Model} 
\label{sec:threat_model}
We consider a centralized IoT authentication architecture comprising a trusted authentication server and a population of resource-constrained IoT devices connected via public, unprotected channels. Each IoT device is equipped with a PUF instance and a read-only memory (ROM) storing cryptographic parameters.
Crucially, the network may comprise \emph{heterogeneous PUF types} from different manufacturers, for example, FPGA boards with Arbiter PUFs, microcontrollers with SRAM PUFs, and Raspberry Pis with DRAM PUFs. Owing to their constrained hardware, IoT devices perform only lightweight local operations. The authentication server is responsible for enrolling devices and verifying their identities based on the trained model. 

The adversary is considered active and capable of observing, intercepting, modifying, or replaying messages transmitted over public communication channels. Beyond channel-level attacks, the adversary may also attempt to fabricate a forgery device equipped with its own PUF and impersonate a legitimate enrolled device. Since the intrinsic manufacturing randomness of a PUF makes it physically unclonable, the forgery device will produce a distinct response pattern even when the same challenge is applied, and our open-set classifier is trained to reject impostors. In the extreme case where the adversary physically captures a legitimate device and attempts to probe or reverse-engineer its PUF circuitry, the underlying physical microstructure will be disrupted, irreversibly damaging the original CRPs due to the tamper-sensitivity of PUF. We assume the adversary cannot compromise the server's private storage or model parameters.

\section{Proposed Method}
\label{sec:openpuf}
An overview of the proposed scheme is shown in Fig.~\ref{fig:overall-open-puf-a}(a). The system relies on training an OpenGAN-based PUF classifier using images derived from PUF responses \textit{solely} collected from legitimate devices. During authentication, PUF images are securely transmitted to the server, where the open-set classifier identifies the device or rejects malicious impersonation attempts. The following subsection extends the image construction method for strong PUFs, whose raw responses do not intrinsically form a 2D bitmap format as do weak PUFs like SRAM and DRAM PUFs.

\subsection{PUF Image Generation}
\label{subsection:generation}
We aim to build a \emph{unified} device-authentication framework for heterogeneous PUF categories, including delay-based and memory-based architectures. In existing classification-based schemes, only memory-based PUFs are adopted because the power-up or decay states inherently form 2D arrays. However, Arbiter PUFs (APUFs)~\cite{sahoo2017multiplexer, zalivaka2018reliable, hemavathy2023arbiter} are a lightweight and widely used strong PUF representative that can be easily implemented in both ASIC and FPGA platforms. The primary challenge is that an APUF produces only a single response bit per challenge, making it incompatible with image-based classification. To address this, we employ a linear feedback shift register (LFSR)~\cite{LFSR2012fpga} to expand an initial random challenge $C$ into a sequence of $N$ challenges. For all PUF types, the final input to the OpenGAN-based classifier is a $W \times H$ grayscale (8 bits per pixel) image. We empirically set $W = H = 50$, requiring $N = 20,000$ response bits to form 2,500 pixels. A detailed ablation study on these dimensions is presented in Sec.~\ref{subsec:ablation}.

During enrollment, a random challenge $C$ is configured as the initial state of the on-device LFSR. When generation is initiated, the LFSR clocks $N$ times to produce expanded challenges $C_{1}, C_{2}, \dots, C_{N}$, which are applied to the strong PUF to yield a response vector $\mathbf{r} = (r_1, r_2, \dots, r_N) \in \{0,1\}^N$. To convert $\mathbf{r}$ into a grayscale image, we group the bits into non-overlapping 8-bit blocks and interpret each as an unsigned integer:
\begin{equation}
v_j = \sum_{b=0}^{7} r_{8(j-1)+b} \cdot 2^b, \quad j = 1,\dots,\frac{N}{8}
\end{equation}
where $v_j \in [0,255]$ represents pixel intensity. The resulting sequence $\{v_j\}$ is reshaped in row-major order into a $W \times H$ image.

For weak PUFs like SRAM and DRAM PUFs, we treat their native 2D cell states as grayscale images, cropping them to the target resolution as necessary. This unified representation enables a solitary open-set classifier to authenticate devices across heterogeneous hardware without PUF-specific adaptations.


\paragraph{Handling noisy and time-varying PUF responses.}
PUF responses are inherently noisy and may drift over time due to environmental fluctuations or aging. Our method trains the classifier to accommodate PUF response instability by repeatedly generating images from the \emph{same} challenge $C$ at different time instants. Each evaluation produces a fresh $N$-bit vector $\mathbf{r}$, resulting in a slightly different grayscale image. These images are treated as distinct training samples, encouraging the CNN to learn stable, device-specific features rather than memorizing a single noise-free template for each device class~\cite{millwood2023dpan, wilde2017large}. We quantify each device's response instability in the public SRAM PUF dataset~\cite{wilde2017large} through the normalized Hamming distance across 101 evaluations per device. Across all 144 devices, this instability averages 5.85\% and can be as high as 26.74\%. The noisy DRAM dataset~\cite{millwood2023dpan} is collected under $20–50^\circ$C and 1.27–1.5 V conditions, deliberately stressing the impact of environmental changes; the strong performance reported in Sec.~\ref{sec:evaluation} under such conditions indicates that the classifier is not just memorizing PUF responses but is robust to response instability.


\paragraph{Security Implications} In contrast to digital twin modeling, classification-based methods do not require modeling the full challenge–response behavior of a PUF~\cite{ruhrmair2013puf, wisiol2022neural, wang2025deep}. Images generated from a single challenge are sufficient to categorize a device within a group. Each device is authenticated based on its unique physical ``fingerprint" generated under a fixed challenge $C$. The initial challenge $C$ and its LFSR-derived variants are not required to be secret; security is derived from the fact that PUF responses depend solely on intrinsic manufacturing variations and cannot be accurately predicted or cloned from the challenge alone. Furthermore, PUF responses are never transmitted in plaintext but are encapsulated within the hybrid-encryption protocols described in Sec.~\ref{sec:openpuf_protocol}. This ensures an adversary \textit{cannot} intercept the raw response bits to mount modeling attacks or predict responses for impersonation. Finally, any attempt to physically probe the PUF will irreversibly disrupt its delicate physical microstructure, rendering the original CRPs invalid and providing inherent protection against hardware tampering.

\subsection{Open-set PUF Classifier}  
\label{sec:classifier}

We adopt OpenGAN~\cite{Opengan2021} to endow the PUF classifier with open-set rejection. Let $\mathcal{D}_{\mathrm{tr}}^L = \{(I_i, y_i)\}_{i=1}^{N_{\mathrm{tr}}}$ denote the training set of PUF images $I_i$ from \emph{legitimate} devices, where $y_i \in \{1,\dots,K\}$ represents the labels of the enrolled devices. The model consists of a $K$-way classifier $\mathbf{M}$, a generator $\mathbf{G}$, and a discriminator $\mathbf{D}$.

\paragraph{Closed-set training}
We first train $\mathbf{M}$ as a standard $K$-way classifier on $\mathcal{D}_{\mathrm{tr}}^L$ using cross-entropy with weight decay:
\begin{equation}
\mathcal{L}_{\mathrm{closed}}
= -\frac{1}{N_{\mathrm{tr}}} \sum_{i=1}^{N_{\mathrm{tr}}} \sum_{c=1}^{K} y_{i,c} \log p_c(I_i)
+ \lambda_{\mathrm{closed}} \!\sum_{j} w_j^2
\label{eq:closed_loss_l2}
\end{equation}
where $p_c(I_i)$ is the predicted probability for class $c$, $y_{i,c}$ is the one-hot encoding of label \(y_i\), and $w_j$ are the parameters of $\mathbf{M}$. After this stage, $\mathbf{M}$ is frozen and serves as a feature extractor for the subsequent open-set modeling phase. 

\paragraph{Open-set feature modeling}
The generator $\mathbf{G}$ maps noise vectors $\mathbf{z}\sim\mathcal{N}(\mathbf{0},\mathbf{I})$ to synthetic open-set features $\tilde{\mathbf{f}} = \mathbf{G}(\mathbf{z})$, while the discriminator $\mathbf{D}$ outputs the probability that a feature belongs to a known device. Importantly, we \emph{eschew} the use of real outlier PUF images during training; instead, $\mathbf{D}$ learns to contrast enrolled features with the synthetic outliers generated by $\mathbf{G}$. This outlier-free approach prevents overfitting to specific impostors and enhances generalization to unseen devices in the open-world environment.

The resulting OpenGAN objective in feature space is
\begin{equation}
\begin{aligned}
\max_{\mathbf{D}} \min_{\mathbf{G}}\ 
& \mathbb{E}_{I \sim \mathcal{D}_{\mathrm{tr}}^L} \big[ \log \mathbf{D}(\phi(I)) \big] \\
& + \lambda_g \, \mathbb{E}_{\mathbf{z} \sim \mathcal{N}(\mathbf{0}, \mathbf{I})} \big[ \log \big( 1 - \mathbf{D}(\mathbf{G}(\mathbf{z}))\big) \big]
\end{aligned}
\end{equation}
where $\phi(I)$ denotes the pre-logit feature extracted by the frozen $K$-way classifier $\mathbf{M}$, and $\lambda_g$ is a hyperparameter balancing the influence of enrolled and synthetic samples.

\paragraph{Inference}
Given a query image $I$, $\mathbf{M}$ outputs a predicted label $\hat{y}$ and a pre-logit feature $f$:
\begin{equation}
\hat{y} = \arg\max_{c} p_c(I), \qquad \mathbf{f} = \phi(I)
\end{equation}
where $\mathbf{D}$ produces an open-set confidence $P_{\mathrm{open}} = \mathbf{D}(\mathbf{f}) \in [0,1]$. The server accepts the claimed device identity if $P_{\mathrm{open}} > \tau$ and $\hat{y}$ matches the claimed ID; otherwise, the query is rejected.

\begin{figure}[t]
    \centering
    \includegraphics[width=1\linewidth]{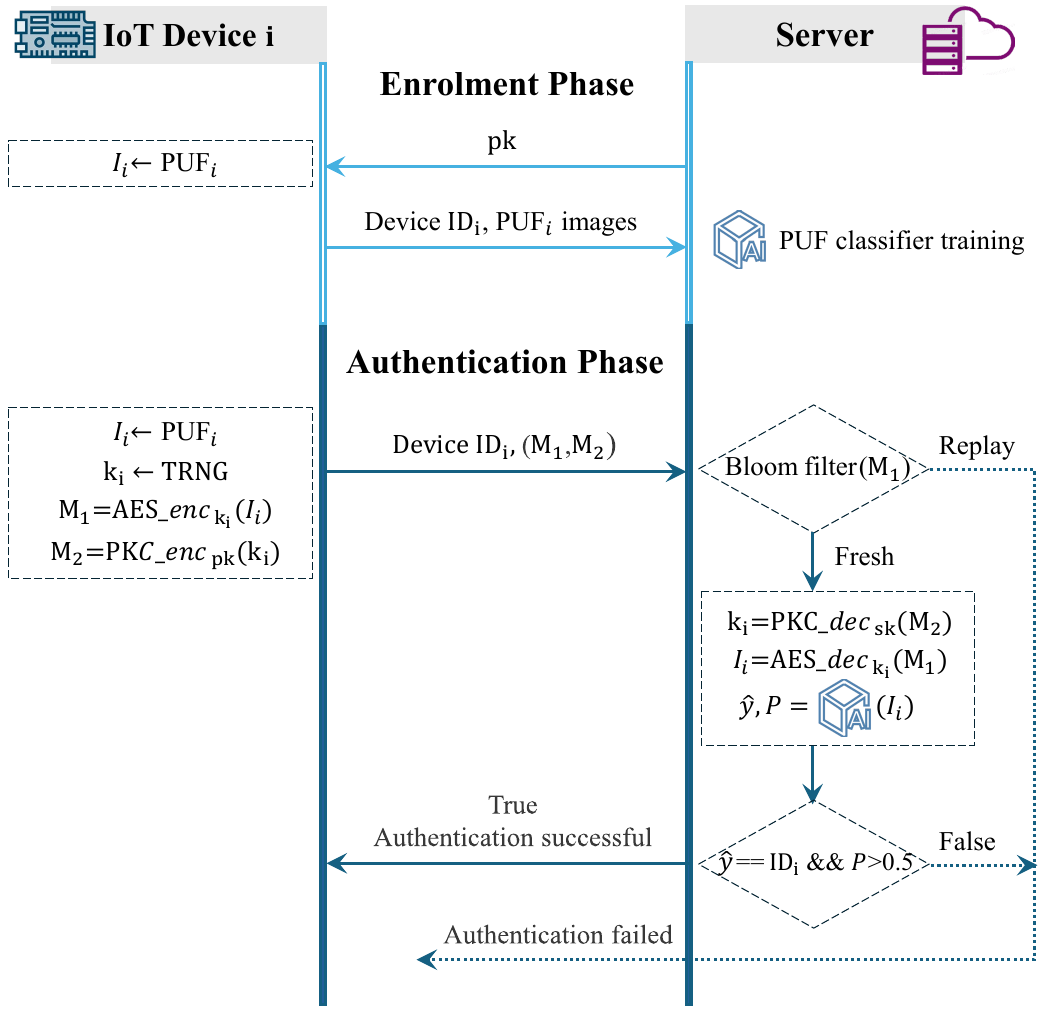}
    \caption{A generic authentication protocol incorporated in our framework.} 
    
    \label{fig:protocol_frame}
    \vspace{-8pt}
\end{figure}

\subsection{A Generic Secure Authentication Protocol}
\label{sec:openpuf_protocol}
To demonstrate the practical feasibility and efficiency of our framework, we integrate it into a generic secure authentication pipeline (see Fig.~\ref{fig:protocol_frame}) employing hybrid encryption and Bloom filter-based replay detection. By securing the transmission process, this design eliminates the critical vulnerabilities to replay and Man-in-The-Middle (MITM) attacks that plague prior approaches~\cite{yue2020dram, millwood2023dpan}, which rely on the plaintext transmission of raw PUF images. Our framework is both protocol- and hardware-agnostic: it can be seamlessly incorporated into standard cryptographic pipelines, such as the Elliptic Curve Integrated Encryption Scheme (ECIES) or neural-network-based protocols~\cite{frank2022memristor, joshi2025error}. Since heterogeneous PUF responses are unified into a standard image representation (Sec.~\ref{subsection:generation}), the entire enrollment and authentication flow operates independently of the underlying PUF architectures.

\paragraph{Enrollment}
The enrollment phase is conducted in a secure environment. The server generates a public-private key pair $(pk, sk)$ using a selected public key cryptography (PKC) algorithm, such as RSA, Elliptic Curve Cryptography, or Post-Quantum Cryptography. The server shares the public key $pk$ with all legitimate IoT devices, which store it in ROM. Each device then applies the enrolled challenge to its embedded PUF module to generate the base PUF images, as described in Sec.~\ref{subsection:generation}. The server collects these images to train the OpenGAN-based classifier described in Sec.~\ref{sec:classifier}. Once trained, the classifier is deployed for field verification.

\paragraph{Authentication}
During authentication, device $i$ regenerates its PUF image $I_i$ by applying the stored challenge $C$. To ensure secure transmission, hybrid encryption is adopted: a fresh session key $k_i$, generated via a true random number generator, encrypts $I_i$ using symmetric encryption (e.g., AES) to produce ciphertext $M_1$. Simultaneously, $k_i$ is encrypted under the server's public key $pk$ to produce $M_2$ for secure key exchange. The device sends its identity $ID_i$ and the messages $(M_1, M_2)$ to the server. Upon receipt, the server first queries its Bloom filter with $M_2$ to detect potential reply attacks. If $M_2$ has been previously seen, the request is immediately rejected. Otherwise, the server decrypts $M_2$ using its private key $sk$ to recover $k_i$, then decrypts $M_1$ to retrieve $I_i$. After a successful authentication, $M_2$ is inserted into the Bloom filter to prevent future replay. Finally, the recovered $I_i$ is processed by the OpenGAN classifier; The device is authenticated only if the open-set confidence $p$ exceeds a threshold $\tau$ and the predicted identity $\hat{y}$ matches the declared $ID_i$. In our prototype, the total data overhead is 2,958 bytes per authentication using a $50\times50$ image with RSA-OAEP and AES-256-GCM.

\section{Evaluation}
\label{sec:evaluation}

\subsection{Datasets and Implementation}
\label{sec:dataset}

\paragraph{\textbf{APUF dataset}}
We implemented 32-stage APUFs on six FPGA boards (Nexys4-DDR, ALINX AXU2CGB-E V2.0, and ALINX AX7Z020B). By instantiating six APUFs per board at different locations, we simulated 36 devices. For each APUF, we fix $C$ to a random 32-bit value and generate 100 images at different time, as described in Sec.~\ref{subsection:generation}. Of these 36 devices (image size: $220 \times 200$), 25 ($A_{1\sim25}$) are designated as legitimate. The remaining 11 APUFs ($A_{26\sim36}$) and 5 DRAM PUFs~\cite{millwood2023dpan} serve as impostors in open-set evaluation.

\paragraph{\textbf{DRAM PUF dataset}}
The noisy DRAM PUF dataset from~\cite{millwood2023dpan} (image size: $220 \times 200$, checkerboard pattern) provides data from five devices measured under varying voltages and temperatures. For closed-set evaluation, all five devices are considered legitimate. For open-set evaluation, we treat three devices as legitimate, reserving one for validation and one for testing as an impostor. To further evaluate outlier rejection, we augment the impostor sets with 52 SRAM ($220 \times 200$) and one additional DRAM device.

\paragraph{\textbf{SRAM PUF dataset}}
The SRAM PUF dataset from~\cite{wilde2017large} contains 144 devices with 101 measurements per device. We select 40 devices as legitimate and the remaining 104 as impostors, split equally (52 each) for validation and testing in open-set experiments.

\paragraph{\textbf{Heterogeneous PUF dataset}}
To test whether the framework can learn a unified feature space across structurally distinct PUF types, we build a mixed PUF dataset combining a subset of APUFs and SRAM PUFs as legitimate devices. This setup specifically tests the framework's ability to generalize across heterogeneous PUF hardware architectures.

\begin{table}[htbp]
\centering
\caption{Dataset configuration (number of devices).}
\label{tab:dataset_configuration}
\renewcommand{\arraystretch}{1}
\begin{threeparttable}
\resizebox{\columnwidth}{!}{
\begin{tabular}{lccc}
\toprule
\textbf{Dataset} & \textbf{Train (legit)} & \textbf{Val (legit / imp.)} & \textbf{Test (legit / imp.)} \\
\midrule
APUF (ours) & 25 & 25 / 6\tnote{(a)} & 25 / 10\tnote{(b)} \\
DRAM PUF closed-set~\cite{millwood2023dpan} & 5 & 5 / 0 & 5 / 0 \\
DRAM PUF open-set~\cite{millwood2023dpan} & 3 & 3 / 53\tnote{(c)} & 3 / 53\tnote{(c)} \\
SRAM PUF~\cite{wilde2017large} & 40 & 40 / 52 & 40 / 52 \\
Heterogeneous PUF & 45 & 45 / 67\tnote{(d)} & 45 / 68\tnote{(e)} \\

\bottomrule
\end{tabular}
}
\small
\begin{tablenotes}\footnotesize
\item \parbox[t]{0.98\columnwidth}{Impostors: (a)~4 APUFs + 2 DRAMs; (b)~7 APUFs + 3 DRAMs; (c)~1 DRAM + 52 SRAMs; (d)~8 APUFs + 59 SRAMs; (e)~8 APUFs + 60 SRAMs. \\ Each legitimate device's images are split into train/val/test sets at a 3:1:1 ratio.}
\end{tablenotes}
\end{threeparttable}
\end{table}

\begin{table*}[!ht]
\caption{Comparison of PUF Classification Methods.}
\label{tab:comparison}
\centering
\begin{threeparttable}
\begin{adjustbox}{width=1\textwidth}
\begin{tabular}{lccccccccc}
\toprule
\textbf{Method} &  \multicolumn{4}{c}{\textbf{Ours}} & \textbf{Reproduced\cite{BCNN2024}} & \multicolumn{1}{c}{\textbf{Mexis et al.\cite{BCNN2024}}} & \textbf{Millwood et al.\cite{millwood2023dpan}} & \textbf{Karimian et al.\cite{yue2020dram}}  & \textbf{Joshi et al.\cite{joshi2025error}} \\
\midrule
\textbf{PUF Type} & Heterogeneous & Arbiter & SRAM & DRAM (noisy) & DRAM (noisy) & DRAM (nominal) & DRAM (noisy) & DRAM (nominal) & SRAM \\ 
\textbf{\# Enrolled Devices} & 45 & 25 & 40 & 3 & 3 & 3 & 3-5 & 3 & 3\\
$\textbf{\textit{Acc}}_{\emph{closed-set}}$ & 100\% & 100\% & 100\% & 100\% & 100\% & 100\% & 92.7\%-98.4\% & 98.6\% & 94.72\%-98.56\% \\
\textbf{False Acceptance Rate} & 0.44\% & 0.35\% & 0.13\% & 0.33\% & 0\% & $0\%^*$ & \ding{55} & \ding{55} & \ding{55} \\
\textbf{False Rejection Rate} & 0.43\% &0\% & 0.62\% & 2.04\% & 10.64\% & $0\%^*$ & \ding{55} & \ding{55} & \ding{55} \\
\textbf{\# Test Samples (\# Outlier Samples)} & 7784 (6859) & 1440 (940) & 6052 (5252) & 5380 (5332) & 5380 (5332) & 172 (43) & 116/154/192 & 108 & 60\\

\textbf{Model Size} & \multicolumn{4}{c}{57.3 MB} & 40.4 MB & unknown & 57MB & unknown & 13.6 MB \\

\textbf{Authentication Time} 
& $\approx 0.57\,\mathrm{s}$ 
& $\approx 0.67\,\mathrm{s}$ 
& $\approx 0.57\,\mathrm{s}$ 
& $\approx 0.64\,\mathrm{s}$ 
& Unknown 
& $\approx 20\,\mathrm{s}$ 
& $5.72$--$7.43\,\mathrm{s}$ 
& Unknown 
& Unknown \\
\textbf{Replay/MITM Attack Resistance} & \ding{51} & \ding{51} & \ding{51}&\ding{51} & \ding{55} & \ding{51} & \ding{55} & \ding{55} & \ding{51} \\
\bottomrule
\end{tabular}
\end{adjustbox}
\begin{tablenotes}
\item Heterogeneous: simultaneously enrolled 20 APUFs and 25 SRAM PUFs;

\item DRAM(noisy): the dataset was collected under various temperature (20°C$\sim$50°C) and voltage settings (1.5V and 1.27V); 
\item DRAM(nominal): the dataset was collected under room temperature and nominal voltage, and is not open-sourced; 
\item[$^*$]: FAR and FRR in \cite{BCNN2024} are evaluated in a tiny dataset with only 43 outlier samples. 
\end{tablenotes}
\end{threeparttable}
\end{table*}

\paragraph{\textbf{Training}}
For all experiments, we use ResNet-18~\cite{he2016deep} as the $K$-way classifier, trained for 10 epochs using the AdamW optimizer~\cite{loshchilov2017fixing} (learning rate $1\times10^{-4}$, weight decay $1\times10^{-3}$) and standard cross-entropy loss. The generator and discriminator, both configured with hidden dimensions $n_{g}=n_{d}=256$ \cite{Opengan2021}, are trained on features extracted from ResNet for 50 epochs with a batch size of 256, using the AdamW optimizer ($\beta_1=0.9, \beta_2=0.999$) with learning rates of $3\times10^{-4}$ and $1.2\times10^{-4}$, respectively, and a weight decay of $1\times10^{-3}$. Binary cross-entropy loss is applied to both models with label smoothing~\cite{szegedy2016rethinking} (real label 0.95, fake label 0.05). Following standard practice for using RGB architectures with grayscale data \cite{rajpurkar2017chexnet, xie2018pre}, we replicate each grayscale PUF image into three channels to form an RGB input, which is normalized before inference.

\paragraph{\textbf{Prototype Implementation}} To demonstrate end-to-end feasibility, we prototype the protocol using a Raspberry Pi 5 as the authentication server and two Ultra96-V2 boards, one acting as a legitimate device and the other as an impostor. The Ultra96-V2 boards, implementing APUFs, run PYNQ 3.0.1 and communicate with the Raspberry Pi over Wi-Fi via the socket package. The trained PUF classifier is exported to ONNX and executed on the Raspberry Pi using ONNX Runtime. For the cryptographic layer, we use RSA for public-key operations and AES for symmetric encryption. We deploy a Bloom filter with 13 hash functions and a 2.4\,MB memory footprint, supporting $10^6$ sessions at a 0.01\% false positive rate (FPR). This capacity is scalable, for instance, a 24\,MB memory can support $10^7$ sessions. Furthermore, the system can be dynamically extended using Scalable Bloom Filters (SBFs) or Bloom Filter Arrays (BFAs)~\cite{almeida2007scalable, wei2010mad2, dayan2023infinifilter} to meet future demand without service interruption. The classifier inference requires 33.5\,GFLOPs, and the end-to-end authentication latency (from protocol initiation to server decision), including encryption, decryption, communication, and inference, is about 0.67\,s. 

\subsection{Performance and Comparison}
Evaluated across the four datasets in Sec. \ref{sec:dataset}, we report: (i) closed-set classification accuracy $Acc_\emph{closed-set}$ for legitimate devices, and (ii) open-set performance via FAR and FRR. As shown in TABLE~\ref{tab:comparison}, our method achieves a 100\% closed-set classification rate, a 0.35\% FAR, and a 0\% FRR for a network of 25 APUF-embedded devices. These results indicate excellent performance in identifying legitimate devices while effectively rejecting impostors. Moreover, our method achieves a 0.13\% FAR and 0.62\% FRR when distinguishing 40 enrolled SRAM PUF instances against more than 5000 unauthorized queries. On the noisy DRAM PUF dataset from~\cite{millwood2023dpan}, we achieve 100\% $\textit{Acc}_{\emph{closed-set}}$, 0.33\% FAR, and 2.04\% FRR, demonstrating robust compatibility with diverse PUF categories. To further test heterogeneous support, we simultaneously enroll 20 APUFs and 25 SRAM PUFs; the system maintains 100\% closed-set accuracy with 0.44\% FAR and 0.43\% FRR, confirming that a single OpenGAN-based classifier can authenticate devices across different hardware types.

We compare our method with other classification-based PUF authentication schemes. CRP database-centric and model-based approaches are not included as baselines, since they represent a fundamentally different design paradigm: each enrolled device requires its own CRP database or predictive model, with storage and maintenance costs that scale linearly with the device population. In contrast, our framework authenticates all devices through a single classifier regardless of PUF type, making direct comparison across these paradigms neither fair nor meaningful.

As summarized in TABLE~\ref{tab:comparison}, three related schemes~\cite{joshi2025error, yue2020dram, millwood2023dpan} were evaluated only in closed-set settings and lack impostor rejection mechanisms. Additionally, schemes~\cite{yue2020dram} and ~\cite{millwood2023dpan} transmit PUF images without cryptographic protection, leaving them vulnerable to replay and MITM attacks. While Mexis et al.~\cite{BCNN2024} provide protocol-level protection against replay/MITM attacks with 100\% closed-set accuracy evaluated using only a single private DRAM PUF dataset acquired in a nominal environment, they require dozens of stochastic forward passes, resulting in a high authentication latency of \textasciitilde20\,s. For a fair comparison, we reproduced their design on the public DRAM dataset from~\cite{millwood2023dpan} (including voltage and temperature variations). Using the Blitz library for 50 epochs and 20 Monte Carlo iterations per query as per~\cite{BCNN2024}, we found that BCNN tends to over-reject enrolled devices on this challenging dataset, yielding a significantly higher FRR (10.64\%) than our OpenGAN-based classifier (FRR 2.04\%, FAR 0.33\%). 

Overall, our method achieves 100\% closed-set accuracy across all configurations, including a heterogeneous configuration, while maintaining low FAR and FRR in open-set evaluation. To our knowledge, this is the first classification-based scheme to demonstrate heterogeneous device authentication. In addition, our prototype achieves an end-to-end authentication time of less than 0.67\,s, approximately $9 \times$ faster than Millwood et al.~\cite{millwood2023dpan} and $30\times$ faster than BCNN~\cite{BCNN2024}.

\subsection{Ablation Studies}
\label{subsec:ablation}
We conduct ablation studies on three key factors. In addition to FRR and FAR, we add AUROC and $F_1$ score to holistically evaluate the trade-off between robustness (rejecting unknown devices) and usability (accepting enrolled devices). 
1) \textbf{PUF Image Size:} TABLE~\ref{tab:sram_size_ablation} indicates that $50\times50$ images offer the best trade-off between accuracy and impostor rejection. Larger images lead to overfitting and poor generalizability, while images that are too small ($25\times25$) lack sufficient discriminative power.
2) \textbf{Discriminator Hidden Dimension $n_{d}$:} TABLE~\ref{tab:nd_performance} shows that $n_{d}=256$ provides the optimal balance between rejection and acceptance. A smaller hidden dimension of 64 yields high error rates due to underfitting, while a larger capacity of 384 tends to over-reject input samples, increasing the FRR.
3) \textbf{Scalability With Enrolled Devices:} TABLE~\ref{tab:sram_scalability} demonstrates that our method maintains high accuracy and low FAR/FRR as the number of enrolled devices increases from 5 to 40, which confirms superior scalability compared to previous methods typically evaluated on 3--5 devices.

\begin{table}[h!]
\centering
\caption{Impact of image size on SRAM PUF dataset (40 devices). Mean ${\pm}$ std over 5 random seeds.}
\label{tab:sram_size_ablation}
\resizebox{1\columnwidth}{!}{
\begin{tabular}{ccccc}
\toprule
Size & FRR (\%) & FAR (\%) & AUROC & F1\\
\midrule
$25\times25$   & $2.40 {\pm} 1.85$     & $1.84 {\pm} 1.15$    & $0.9938 {\pm} 0.0102$ & $0.9320 {\pm} 0.0409$ \\
$50\times50$   & \bm{$1.20{\pm}0.68$}  &\bm{$0.05{\pm}0.05$} & \bm{$0.9993{\pm}0.0003$} & \bm{$0.9927{\pm}0.0038$} \\
$100\times100$ & $6.90 {\pm} 3.75$     & $0.51 {\pm} 0.86$    & $0.9986 {\pm} 0.0005$ & $0.9508 {\pm} 0.0210$ \\
$200\times200$ & $18.47 {\pm} 3.76$    & $4.18 {\pm} 3.16$    & $0.8963 {\pm} 0.1814$ & $0.5620 {\pm} 0.3892$ \\
\bottomrule
\end{tabular}}
\end{table}

\begin{table}[h!]
\centering
\caption{Impact of discriminator hidden dimension ($n_{\text{d}}$) on SRAM PUF dataset (image size: $50\times50$). Mean ${\pm}$ std over 5 random seeds.}
\label{tab:nd_performance}
\renewcommand{\arraystretch}{1}
\resizebox{1\columnwidth}{!}{
\begin{tabular}{ccccc}
\toprule
$n_{\text{d}}$ & FRR (\%) & FAR (\%) & AUROC & F1 \\
\midrule
64  & $8.00 {\pm} 2.20$   & $6.24 {\pm} 3.99$   & $0.9625 {\pm} 0.0186$ & $0.8176 {\pm} 0.0843$ \\
128 & \bm{$1.12 {\pm} 0.74$}   & $0.46 {\pm} 0.43$   & $0.9993 {\pm} 0.0008$ & $0.9796 {\pm} 0.0109$ \\
$256$ & $1.2{\pm}0.68$ & $0.05{\pm}0.05$  & \bm{$0.9993{\pm}0.0003$} & \bm{$0.9927{\pm}0.0038$} \\
384 & $18.33\pm14.18$   & \bm{$0.00\pm0.00$}            & $0.9896\pm0.0190$    & $0.8936 {\pm}0.1040$ \\
\bottomrule
\end{tabular}}
\end{table}

\begin{table}[h!]
\centering
\caption{Scalability on SRAM PUF devices (image size: $50\times50$). Mean ${\pm}$ std over 5 random seeds.}
\label{tab:sram_scalability}
\renewcommand{\arraystretch}{1}
\resizebox{1\columnwidth}{!}{
\begin{tabular}{ccccc}
\toprule
\#devices & FRR (\%) & FAR (\%) & AUROC & F1 \\
\midrule
5   & $0.20 {\pm} 0.45$ & $0.38 {\pm} 0.22$ & $1.0000 {\pm} 0.0000$ & $0.9085 {\pm} 0.0477$ \\
10  & $0.90 {\pm} 0.65$ & $0.31 {\pm} 0.60$ & $0.9999 {\pm} 0.0001$ & $0.9602 {\pm} 0.0657$ \\
20  & $0.70 {\pm} 0.33$ & $0.01 {\pm} 0.01$ & $0.9985 {\pm} 0.0007$ & $0.9962 {\pm} 0.0015$ \\
30  & $0.73 {\pm} 0.49$ & $0.03 {\pm} 0.06$ & $0.9992 {\pm} 0.0004$ & $0.9950 {\pm} 0.0037$ \\
40  &$1.20{\pm}0.68$  &$0.05{\pm}0.05$ & $0.9993{\pm}0.0003$ & $0.9927{\pm}0.0038$ \\

\bottomrule
\end{tabular}}
\end{table}

\section{Conclusion}
This paper proposes a novel PUF-based authentication scheme that supports the accurate verification of legitimate devices and robust rejection of unknown ones. Leveraging the OpenGAN technique, our open-set PUF authentication framework achieves strong performance on three single-type PUF datasets, Arbiter, SRAM~\cite{wilde2017large}, and noisy DRAM~\cite{millwood2023dpan}, while demonstrating superior compatibility with heterogeneous PUF types. Across four experimental settings, the proposed method attains 100\% closed-set accuracy with low open-set error rates: 0.35\% FAR / 0\% FRR on Arbiter PUFs, 0.13\% FAR / 0.62\% FRR on SRAM PUFs, 0.33\% FAR / 2.04\% FRR on highly noisy DRAM PUFs, and 0.44\% FAR and 0.43\% FRR on heterogeneous PUFs. Additionally, we incorporate a generic authentication protocol that combines hybrid encryption with a Bloom filter to defend against replay and MITM attacks. Compared with existing PUF classification schemes, our method offers superior open-set authentication performance, efficient inference, improved scalability, and heterogeneous compatibility, establishing it as a practical solution for securing heterogeneous IoT networks.

\bibliographystyle{IEEEtran}
\bibliography{references}

\end{document}